\documentclass[letterpaper,english,reprint, aps]{revtex4-1}
\usepackage[LGR,T1]{fontenc}
\usepackage[latin9]{inputenc}
\usepackage{color}
\usepackage{textcomp}
\usepackage{amsmath}
\usepackage{amssymb}
\usepackage{graphicx}
\usepackage{esint}

\makeatletter

\pdfpageheight\paperheight
\pdfpagewidth\paperwidth

\DeclareRobustCommand{\greektext}{%
  \fontencoding{LGR}\selectfont\def\encodingdefault{LGR}}
\DeclareRobustCommand{\textgreek}[1]{\leavevmode{\greektext #1}}
\ProvideTextCommand{\~}{LGR}[1]{\char126#1}

\makeatother

\usepackage{babel}
\begin{document}

\preprint{$\enspace$}

\title{A quantum retrograde canon:\\
Complete population transfer in $n^{2}$-state systems}

\author{Alon Padan}

\author{Haim Suchowski}
\email{haimsu@post.tau.ac.il}

\affiliation{Raymond and Beverly Sackler School of Physics and Astronomy, Tel
Aviv University, Ramat Aviv, Tel Aviv 69978, Israel}

\date{\today}
\begin{abstract}
\textcolor{black}{We present a novel approach for analytically reducing
a family of time-dependent multi-state quantum control problems to
two-state systems. The presented method translates between $SU(2)\text{\texttimes}SU(2)$
related $n^{2}$-state systems and two-state systems, such that the
former undergo complete population transfer (CPT) if and only if the
latter reach specific states. For even $n$, the method translates
any two-state CPT scheme to a family of CPT schemes in $n^{2}$-state
systems. In particular, facilitating CPT in a four-state system via
real time-dependent nearest-neighbors couplings is reduced to facilitating
CPT in a two-level system. Furthermore, we show that the method can
be used for operator control, and provide conditions for producing
several universal gates for quantum computation as an example. In
addition, we indicate a basis for utilizing the method in optimal
control problems.}
\end{abstract}
\maketitle
\textcolor{black}{Multi-state quantum systems are in the frontier
of quantum control research, holding high potential for reducing computation
complexity \citep{Lanyon_2008}, and strengthening communication security
\citep{PhysRevLett.88.127902} in quantum information technologies
\citep{nielsenm.a.i.l.chuang2002}. A central topic in multi-state
control is complete population transfer (CPT), where a system is transferred
between orthogonal states. This issue is a fundamental task in applications
such as state preparation and entanglement transfer \citep{PhysRevLett.82.1971}. }

\textcolor{black}{The simplest and most studied quantum control problems
deal with two-state systems \citep{allenl.j.h.eberly1975},\citep{1751-8121-41-15-155309}.
As we move to multiple states, synthesis and analysis of control schemes
become increasingly difficult. Hence, multi-state control problems
are often approached by some method of reduction to one or more two-state
problems \citep{driftsromanodelassandro},\citep{PhysRevA.84.063413}.
One such method is adiabatic elimination \citep{PhysRevA.60.3081},
in which the system\textquoteright s dynamics are approximated by
a two-state system through the elimination of irrelevant, non-resonantly
coupled states. While this approach is often adequate, its applicability
is conditioned on parameters\textquoteright{} range and its results
may suffer significant inaccuracies under multiple eliminations \citep{Paulisch2014}.
Another reduction method \citep{Hioe:87} provides analytic solutions
to $N$-level systems with $SU(2)$ dynamic symmetry in terms of the
their fundamental two-state representation. While this approach is
exact, it is limited to systems with $su(2)$ dynamical Lie algebra,
and thus have only three degrees of freedom at each moment of time.}

\textcolor{black}{In this paper, we present the }\textit{\textcolor{black}{quantum
retrograde canon}}\textcolor{black}{, a novel method for analytically
reducing time-dependent multi-state quantum control problems to two-state
systems. The reduction method is based on an exact translation between
two-state systems and time-dependent $n^{2}$-state Hamiltonians with
$su(2)\oplus su(2)$ dynamical Lie algebra. The principle idea underlying
the translation resembles the technique of the retrograde canon in
music, where a musical line is played simultaneously with another
copy of it, inverted in time (see ``Crab Canon'' by Bach \citep{crabcanon}).
Analogously, the quantum retrograde canon }maps a one-qubit scheme
to a two-qubit ``retrograde canon\textcolor{black}{'' - taking one
qubit through the dynamics of the original system and the other qubit
through the same dynamics, inverted in time. Given an appropriate
choice of basis, the two-qubit system undergoes CPT if and only if
the one-qubit system reaches a specific state. Since the mapping is
invertible, one can apply the method the other way around and translate
two-qubit schemes, which have six degrees of freedom at each moment,
to two-state schemes, which have only three degrees of freedom at
each moment. By using higher order representation of the two-qubit
system, and an appropriate choice of basis, one gets a translation
method between two-state schemes and $SU(2)\text{\texttimes}SU(2)$
controlled $n^{2}$-state CPT schemes. In particular, for even $n$,
the method translates any two-state CPT scheme to CPT schemes in $n^{2}$-state
systems.}

\textcolor{black}{The paper is structured as follows: In section I,
we define the retrograde canon. In section II, we illustrate a way
in which the retrograde canon reduces four-level CPT problems to two-level
CPT problems. In section III, we formulate and prove the fundamental
observation underlying the reduction method. In section IV, we state
a more general claim regarding four-level CPT problems. In section
V, we discuss two concrete examples of the method's application. In
section VI, we indicate the use of the method for optimal and operator
control problems. Lastly, in section VII, we consider generalizations
of the method, including application to $n^{2}$-state systems through
higher order representations of the $SU(2)$ group. In section VIII,
we summarize. Various technical issues are discussed in the appendices.}

\section{basic definitions }

We begin with defining the key notions of the retrograde canon. Let
$H(t)$ be a time-dependent two state system Hamiltonian and $U(t)$
be the propagator generated by \emph{$H$}, i.e., $U$ is the solution
of Schrödinger's operator equation $\dot{U}(t)=-iH(t)U(t)$ with the
initial condition $U(0)=I_{2}$. We define the \emph{Retrograde }of
$H$, starting from time $T$>\emph{ }0, by 
\begin{equation}
H^{R}(t)\equiv-H(T-t)\label{eq: definition of retrograde Hamiltonian}
\end{equation}

It can be verified by differentiation that $U^{R},$ the propagator
generated by $H^{R}$, satisfies:
\begin{equation}
U^{R}(t)=U(T-t)U(T)^{-1}\label{eq:1the retrograde propagator}
\end{equation}

$U^{R}(t)$ takes states backwards in time along the path they trace
when acted on by $U(t)$. That is, suppose that $\ensuremath{|f\text{\textrangle}}\equiv U(T)|i\text{\textrangle}$
- i.e., $|f\text{\textrangle}$ is the state to which the initial
state $|i\text{\textrangle}$ evolves to in the original system -
then 
\begin{equation}
\ensuremath{U^{R}(t)|f\text{\textrangle}}=U(T-t)|i\text{\textrangle}\label{eq:Ur of f equals i-1}
\end{equation}
 In what follows we will be interested in the the effect of acting
simultaneously with both the original Hamiltonian and its retrograde.
Thus, we define the \emph{Retrograde canon Hamiltonian}, $H^{RC}$,
by
\begin{equation}
H^{RC}(t)\equiv H^{R}(t)\text{\ensuremath{\otimes}\ensuremath{\ensuremath{I_{2}}+\ensuremath{I_{2}\otimes H}(t)}}\label{eq: retrograde canon hamiltonian definition}
\end{equation}
 $H^{RC}$ acts on one qubit with the original time-dependent Hamiltonian,
and on another qubit with its retrograde, starting from time $T>0$.
clearly, $U^{RC}$, the \emph{Retrograde Canon Propagator} generated
by $H^{RC}$, satisfies: 
\begin{equation}
U^{RC}(t)=U^{R}(t)\otimes U(t)\label{eq:2retrograde canon operator-1}
\end{equation}

The retrograde canon is a one-to-one mapping from a $su(2)$ two-state
time-dependent Hamiltonian in the time interval $[0,\,T]$ and a $su(2)\oplus su(2)$
four-state time-dependent Hamiltonian in the time interval $[0,\:T/2]$.
Accordingly, any time-dependent four-state Hamiltonian of the form
$\tilde{H}(t)=H^{b}(t)\text{\ensuremath{\otimes}\ensuremath{\ensuremath{I_{2}}+\ensuremath{I_{2}\otimes H^{a}}(t)}}$,
where $H^{b}(t),H^{a}(t)$ are two-state Hamiltonians defined for
$t\in[0,\:T/2]$, can be regarded as the retrograde canon of a unique
two-state Hamiltonian, $H$, defined for $t\in[0,T]$ through 
\begin{equation}
H(t)=\begin{cases}
H^{a}(t) & 0\leq t\leq T/2\\
-H^{b}(T-t) & T/2\leq t\leq T
\end{cases}\label{eq: the inverse of the retrograde canon}
\end{equation}
Thus, the 6 degrees of freedom of the four-level momentary Hamiltonian
$\tilde{H}(s)$ are mapped to 3+3 degrees of freedom of the two two-level
momentary Hamiltonians, $H(s)$ and $H(T-s)$.

\section{AN APPLICATION OF THE METHOD}

Let's present a relatively general case of reducing CPT problems from
four-state systems to two-state systems through the retrograde canon.
This is not the most general four-level application of the method,
but it is general enough to understand the idea. Consider the following
four-level Hamiltonian: 
\begin{equation}
\mathcal{H}(t)=\begin{pmatrix}0 & A(t) & iE(t) & D(t)\\
A(t) & 0 & B(t) & iF(t)\\
-iE(t) & B(t) & 0 & C(t)\\
D(t) & -iF(t) & C(t) & 0
\end{pmatrix}\label{eq: two four level Hamiltonian}
\end{equation}

where $A(t),...,F(t)$ are six arbitrary real functions of time. Suppose
we are interested in the conditions under which $\mathcal{H}(t)$
facilitates CPT from $\Psi_{1}=(1,0,0,0)^{\intercal}$ to $\Psi_{3}=(0,0,1,0)^{\intercal}$
at some time $\tau>0$, i.e., the conditions under which $1=|\text{\textlangle}\Psi_{3},\mathcal{U}(\tau)\Psi_{1}\text{\textrangle}|$,
where $\mathcal{U}(t)$ is the propagator generated by $\mathcal{H}(t)$.
Special cases of this problem are encountered in the literature \citep{PhysRevA.60.3081}\citep{PythagoreanPhysRevA84013414}\citep{hiddenqubit}
and in applications. We will show that the through the retrograde
canon we reduce this question to the question whether $e_{1}=(1,0)^{\intercal}$
evolves to $\pm e_{2}=\pm(0,1)^{\intercal}$ at time $T=2\tau$ in
a two-state system governed by the Hamiltonian $H(t)=x(t)\sigma_{x}+y(t)\sigma_{y}+z(t)\sigma_{z}$,
i.e., 
\begin{equation}
H(t)=\begin{pmatrix}z(t) & x(t)+iy(t)\\
x(t)-iy(t) & -z(t)
\end{pmatrix}\label{eq: the two level Hamiltonian}
\end{equation}
where
\[
\begin{array}{c}
x(t)\equiv-\frac{1}{2}\begin{cases}
C(t)-A(t) & 0\leq t\leq T/2\\
C(T-t)+A(T-t) & T/2\leq t\leq T
\end{cases}\\
y(t)\equiv\frac{1}{2}\begin{cases}
E(t)+F(t) & 0\leq t\leq T/2\\
E(T-t)-F(T-t) & T/2\leq t\leq T
\end{cases}\\
z(t)\equiv\frac{1}{2}\begin{cases}
B(t)+D(t) & 0\leq t\leq T/2\\
B(T-t)-D(T-t) & T/2\leq t\leq T
\end{cases}
\end{array}
\]

The translation between the two-level Hamiltonian of eq. \ref{eq: the two level Hamiltonian}
and the four-level Hamiltonian of eq. \ref{eq: two four level Hamiltonian}
is done in two stages, illustrated schematically in Figure 1: In the
first stage, we change basis and define $\tilde{H}(t)\equiv W^{\text{\dag}}\mathcal{H}(t)W$,
where 
\begin{equation}
W\equiv\frac{1}{\sqrt{2}}\begin{pmatrix}1 & 0 & 0 & 1\\
0 & 1 & -1 & 0\\
0 & 1 & 1 & 0\\
1 & 0 & 0 & -1
\end{pmatrix}\label{eq: bell similarity transformation}
\end{equation}

In the second stage, we apply eq. \ref{eq: the inverse of the retrograde canon}
to $\tilde{H}(t)$ and get the two-level $H(t)$. That is, we consider
$\tilde{H}(t)$ as the retrograde canon Hamiltonian of some $H(t)$,
and solve for $H(t)$.

\begin{figure*}[t]
\includegraphics[scale=0.75]{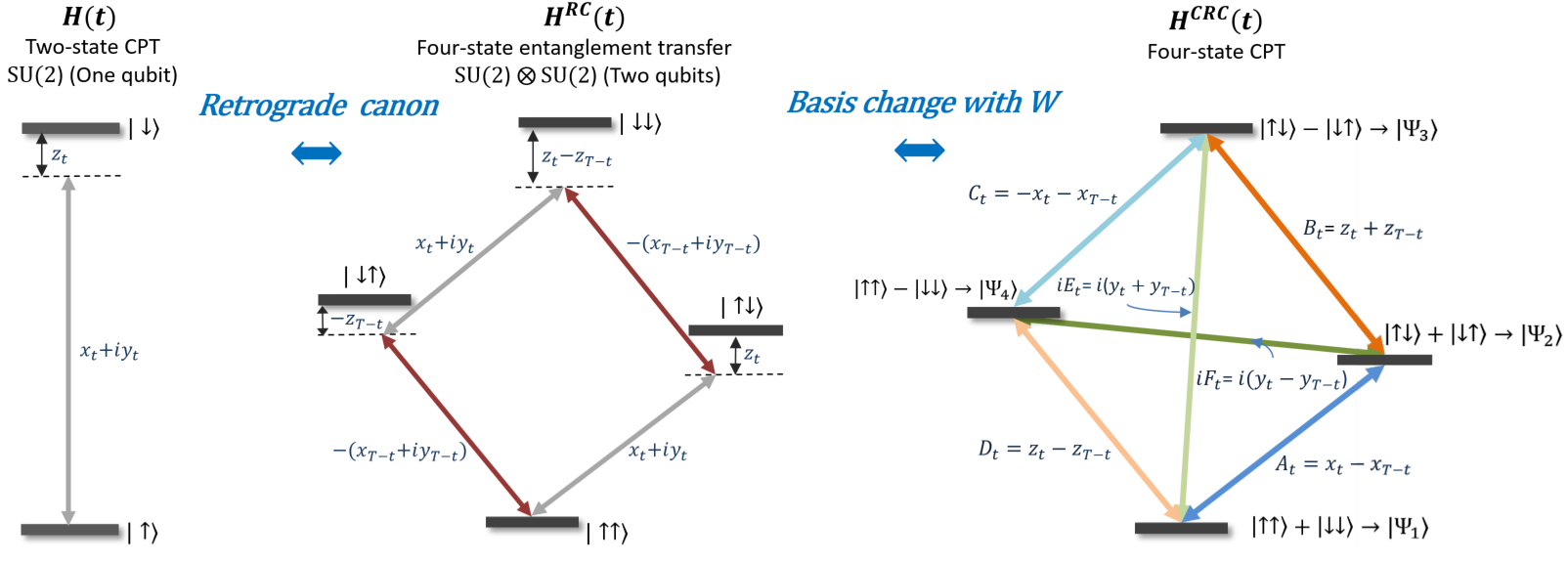}

\caption{Couplings diagrams illustrating the translation method - Going opposite
to the direction of the main text, we start from the two-state system
(on the left, with lower indices indicating time). Applying the retrograde
canon, defined in eq. \ref{eq: retrograde canon hamiltonian definition},
we get a four-state system of two independent qubits (in the middle).
Switching basis using the unitary matrix $W$ defined in eq. \ref{eq: bell similarity transformation},
provides a four-state Hamiltonian of the form given in eq. \ref{eq: two four level Hamiltonian}
(on the right). We show below that the right system undergoes CPT
at time $\tau$ if and only if the left system evolves from $|\downarrow\rangle$
to $\pm|\downarrow\rangle$ at time $2\tau$. }
\end{figure*}

The claim that $\mathcal{H}(t)$ facilitates CPT from $\Psi_{1}=(1,0,0,0)^{\intercal}$
to $\Psi_{3}=(0,0,1,0)^{\intercal}$ at time $\tau$ if and only if
$H(t)$ takes $e_{1}=(1,0)^{\intercal}$ to $\pm e_{2}=\pm(0,1)^{\intercal}$
at time $T=2\tau$ is a special case of the claim we formulate and
prove below. 

\section{the central observation}

Prior to formulating the general claim for the case of four-state
systems, from which the example above follows as a a special case,
let us state and prove the observation which provides the basis for
using the retrograde canon to reduce multi-state CPT problems to two-state
systems:
\begin{equation}
U(T)|\uparrow\rangle=|\downarrow\rangle\Leftrightarrow U^{RC}(\tau)(|\downarrow\downarrow\rangle+|\uparrow\uparrow\rangle)=|\uparrow\downarrow\rangle-|\downarrow\uparrow\rangle\label{eq: CPT iff entangelemnt transfer}
\end{equation}
where $|\uparrow\rangle\equiv(1,0)^{\intercal}$ and $|\downarrow\rangle\equiv e^{i\alpha}(0,1)^{\intercal}$
for some $\alpha\in[0,2\pi)$. Eq. \ref{eq: CPT iff entangelemnt transfer}
states that a two-state Hamiltonian, $H$, facilitates CPT at time
$T$ if and only if $H^{RC}$, facilitates a specific 'entanglement
transfer' at time $\tau\equiv T/2$. We emphasize that the path of
$U(t)$ up to the point $t=T$ can be completely arbitrary. In particular,
$U(\tau)$, which appears in $U^{RC}(\tau)=U(\tau)\otimes U^{R}(\tau)$,
can be any $SU(2)$ matrix. 

To prove the $\Rightarrow$ direction of eq. \ref{eq: CPT iff entangelemnt transfer}
we first note that eq. \ref{eq:Ur of f equals i-1} tells us that
for $\tau=T/2$, 
\begin{equation}
\ensuremath{U^{R}(\tau)|f\text{\textrangle}}=U(\tau)|i\text{\textrangle}\label{eq:Ur of f equals i}
\end{equation}
 Now, assume that indeed $U(T)|\uparrow\rangle=|\downarrow\rangle$.
Since $\ensuremath{U(\tau)}$ is unitary, it transfers orthogonal
states to orthogonal states. Thus, it is sensible to mark $U(\tau)|\uparrow\rangle\equiv|\tilde{\uparrow}\rangle$
and $U(\tau)|\downarrow\rangle\equiv|\tilde{\downarrow}\rangle$.
From eq. \ref{eq:Ur of f equals i} we get that $U^{R}(\tau)|\downarrow\rangle\text{=|\ensuremath{\tilde{\uparrow}\rangle}}$.
Moreover, from the fact that $U(T)^{-1}$ is a $\pi$-rotation in
a spin-half representation, and hence $U(T)^{-1}U(T)^{-1}=-I_{2}$,
it follows that (marking $U_{t}\equiv U(t)$)
\begin{equation}
\begin{array}{c}
U^{R}(\tau)|\uparrow\rangle=\ensuremath{U_{\tau}}U_{T}^{-1}|\uparrow\rangle\text{\ensuremath{=\ensuremath{U_{\tau}}\ensuremath{U_{T}}U_{T}^{-1}U_{T}^{-1}|\uparrow\rangle}}\\
=-\ensuremath{U_{\tau}}\ensuremath{U_{T}}|\uparrow\text{\ensuremath{\rangle}=\ensuremath{-\ensuremath{U_{\tau}}|\downarrow\rangle}}
\end{array}\label{eq: why the minus sign}
\end{equation}
and therefore $U^{R}(\tau)|\uparrow\rangle=-|\tilde{\downarrow}\rangle$
. Hence,
\begin{equation}
\begin{array}{c}
U^{RC}(\tau)(|\downarrow\downarrow\rangle+|\uparrow\uparrow\rangle)=\\
\ensuremath{=U_{\tau}^{R}}|\downarrow\rangle\otimes U_{\tau}|\downarrow\rangle+\ensuremath{U_{\tau}^{R}}|\uparrow\rangle\otimes U_{\tau}|\uparrow\rangle\\
=|\tilde{\uparrow}\tilde{\downarrow}\rangle-|\tilde{\downarrow}\tilde{\uparrow}\rangle=|\uparrow\downarrow\rangle-|\downarrow\uparrow\rangle
\end{array}\label{eq: first claim}
\end{equation}

where the last step follows from the Clebsh-Jordan fact that $|\uparrow\downarrow\rangle-|\downarrow\uparrow\rangle$
is a $SU(2)$ scalar. 

To prove the $\Leftarrow$ direction of eq. \ref{eq: CPT iff entangelemnt transfer}
we utilize a slightly different perspective, which will also be useful
for the generalizations of the retrograde canon presented below.  We
assume that $U^{RC}(\tau)(|\downarrow\downarrow\rangle+|\uparrow\uparrow\rangle)=|\uparrow\downarrow\rangle-|\downarrow\uparrow\rangle$
and need to prove that $U(T)|\uparrow\rangle=|\downarrow\rangle$.
For the proof we shall use two simple facts: the first is that for
any three 2-by-2 matrices, $A,\,B,\,m$, 
\begin{equation}
(A\otimes B)F(m)=F(AmB^{\intercal})\label{eq:double rotation formula}
\end{equation}

where $F$ simply flattens matrices, i.e., $F\left(\begin{pmatrix}a & b\\
c & d
\end{pmatrix}\right)=(a,b,c,d)^{\intercal}.$ Eq. \ref{eq:double rotation formula} may be verified by direct calculation.
The second, is that for $Y\equiv e^{+i\frac{\pi}{2}\sigma_{y}}=\begin{pmatrix}0 & 1\\
-1 & 0
\end{pmatrix}$ ($\sigma_{i}$ are the Pauli matrices) and any $u\in SU(2)$, 
\begin{equation}
uYu^{\intercal}=Y\label{eq: the defining equation of symplectic}
\end{equation}

Eq. \ref{eq: the defining equation of symplectic} is in fact equivalent
to the claim that $|\downarrow\uparrow\rangle-|\uparrow\downarrow\rangle$
is a $SU(2)$ scalar - the equivalence can be seen through eq. \ref{eq:double rotation formula}
and the observation that $|\uparrow\downarrow\rangle-|\downarrow\uparrow\rangle\rightarrow e^{i\alpha}F(Y)$
in the basis $\{|\uparrow\uparrow\rangle,|\uparrow\uparrow\rangle,|\uparrow\uparrow\rangle,|\uparrow\uparrow\rangle\}$
. 

To proceed, we define $R\equiv e^{-i\alpha\sigma_{z}}$ and note that
in the above basis $|\downarrow\downarrow\rangle+|\uparrow\uparrow\rangle\rightarrow F\left(\begin{pmatrix}1 & 0\\
0 & e^{2i\alpha}
\end{pmatrix}\right)=e^{i\alpha}F(R)$. Now, it follows that the assumption is equivalent to 
\begin{equation}
\begin{array}{c}
Y=F^{-1}((U_{\tau}^{R}\otimes U_{\tau})F(R))=U_{\tau}^{R}RU_{\tau}^{\intercal}=\\
=U_{\tau}^{R}RY^{-1}U_{\tau}^{-1}U_{\tau}YU_{\tau}^{\intercal}=U_{\tau}^{R}RY^{-1}U_{\tau}^{-1}Y
\end{array}\label{eq: derivation}
\end{equation}
where we used eq. \ref{eq:double rotation formula} in the second
equality, inserted $I_{2}=Y^{-1}U_{\tau}^{-1}U_{\tau}Y$ in the third
equality, and used eq. \ref{eq: the defining equation of symplectic}
in the fourth equality. Multiplying eq. \ref{eq: derivation} by $Y^{-1}$
we get that $U_{\tau}^{R}RY^{-1}U_{\tau}^{-1}=I_{2}$ and therefore
$(U_{\tau}^{R})^{-1}=RY^{-1}U_{\tau}^{-1}$. So finally we reach our
goal:

\[
\begin{array}{c}
U(T)|\uparrow\rangle=(U_{\tau}^{R})^{-1}U_{\tau}|\uparrow\rangle=RY^{-1}U_{\tau}^{-1}U_{\tau}|\uparrow\rangle=\\
=-RY|\uparrow\rangle=R(0,1)^{\intercal}=e^{i\alpha}(0,1)^{\intercal}=|\downarrow\rangle
\end{array}
\]
where the first equality can be easily derived from eq. \ref{eq: the inverse of the retrograde canon}. 

\section{cpt in four-state systems}

Eq. \ref{eq: CPT iff entangelemnt transfer} describes a correspondence
between two-state CPT and four-state entanglement transfer of $|\downarrow\downarrow\rangle+|\uparrow\uparrow\rangle$
to $|\downarrow\uparrow\rangle-|\uparrow\downarrow\rangle$. It takes
just a few small steps to connect this entanglement transfer to four-state
CPT between standard basis vectors.

Marking $w_{1}\equiv\frac{1}{\sqrt{2}}(|\downarrow\downarrow\rangle+|\uparrow\uparrow\rangle)$
and $w_{3}\equiv\frac{1}{\sqrt{2}}(|\downarrow\uparrow\rangle-|\uparrow\downarrow\rangle)$,
we first note that 
\begin{equation}
U(T)|\uparrow\rangle=\pm|\downarrow\rangle\Leftrightarrow U^{RC}(\tau)w_{1}=\pm w_{3}\label{eq: trivial generalization of CPT iff entanglement}
\end{equation}
 is a trivial generalization of eq. \ref{eq: CPT iff entangelemnt transfer}.
Next, we note that in the fundamental representation of $SU(2)\times SU(2)$
(and in \textit{even} higher order representations), $e^{i\phi_{1}}w_{1}$
and $e^{i\phi_{3}}w_{3}$ are orthonormal for any $\phi_{1},\phi_{3}\in[0,2\pi)$.
They can therefore be completed to an orthonormal basis of $\mathbb{C}^{4}$
with some vectors $w_{2},w_{4}\in\mathbb{C}^{4}$. This basis determines
a unitary \textit{conjugating matrix} $W\equiv(e^{i\phi_{1}}w_{1},w_{2},e^{i\phi_{3}}w_{3},w_{4})\in M_{4}[\mathbb{\boldsymbol{\mathbb{{C}}}}]$
which we use to define
\begin{equation}
\mathcal{H}^{CRC}(t)\text{\ensuremath{\equiv}}W^{\text{\dag}}H^{RC}(t)W\label{eq: Hcrc definition}
\end{equation}

Suppose now that $\mathcal{H}^{CRC}(t)$ facilitates CPT at time $\tau$,
i.e., that $\text{|\textlangle}\Psi_{3},\mathcal{U}^{CRC}(\tau)\Psi_{1}\rangle|=1$
where $\mathcal{U}^{CRC}$ is the propagator generated by $\mathcal{H}^{CRC}$
and $\Psi_{i}$ are the standard basis vectors. It follows that $U^{RC}(\tau)w_{1}=e^{i\beta}w_{3}$
for some $\beta\in[0,2\pi)$. Then, by repeating the steps of eq.
\ref{eq: derivation}, we would conclude that $e^{i\beta}=U_{\tau}^{R}RY^{-1}U_{\tau}^{-1}$
. Since $R,Y$, $U_{\tau}$ and $U_{\tau}^{R}$ are elements in $SU(2)$,
then so is $e^{i\beta}$ and therefore $e^{i\beta}=\pm1$. Hence,
together with eq. \ref{eq: trivial generalization of CPT iff entanglement}
we conclude that
\begin{equation}
U(T)|\uparrow\rangle=\pm|\downarrow\rangle\Leftrightarrow|\text{\textlangle}\Psi_{3},\mathcal{U}^{CRC}(\tau)\Psi_{1}\text{\textrangle}|=1\label{eq: CPT iff CPT in four}
\end{equation}
i.e., $\mathcal{H}^{CRC}$ facilitates CPT at time $\tau$ from $\text{\ensuremath{\Psi_{1}}}$
to $\text{\ensuremath{\Psi_{3}}}$ if and only if $U(T)$ evolves
$|\uparrow\rangle$ to $\pm|\downarrow\rangle$ at time $T=2\tau$.
The claim illustrated in Section II consists in choosing a specific
$W$. A general form of possible conjugating matrices $W$ is presented
in Appendix A, where we also present the general form of the resulting
$\mathcal{H}^{CRC}$. We note that the diagonal couplings of $\mathcal{H}^{CRC}$
are always zero and that with an appropriate choice of $W$ all six
above-diagonal entries of $\mathcal{H}^{CRC}(t)$ can be set imaginary,
yet no more than four can be set real. 

The general form of $\mathcal{H}^{CRC}$ shows which four-state Hamiltonians
can be translated by the retrograde canon method to a two-level system.
For these Hamiltonians, the question of whether they performs CPT
at time $\tau$ be reduced to a question regarding the state of a
two-state system in time $2\tau$. 

\section{examples}

We now present two concrete examples of the method's application.
To better see what's going on in these examples we shall visualize
the evolution of the propagators as a path in the unit 3-ball, $B=\{x\in\mathbb{\mathbb{R}}^{3}|\|x\|\text{\ensuremath{\le}}1\}$.
We quickly review how we do that: Any element $u\in SU(2)$ corresponds
to a rotation by an angle $\phi\in[0,4\pi)$ around an axis $\hat{r}\in\mathbb{R}^{3}\enspace(|\hat{r}|=1)$
according to $u=\hat{R}_{r}(\phi)\equiv e^{i\phi\hat{r}\cdot\vec{J}^{(2)}}$,
where $\vec{J}^{(}{}^{2}{}^{)}\equiv(J_{x}^{(}{}^{2}{}^{)},J_{y}^{(}{}^{2}{}^{)},J_{z}^{(}{}^{2}{}^{)})$
)$\equiv\frac{1}{2}(\sigma_{x},\sigma_{y},\sigma_{z})$. A path in
$SU(2)$ can thus be projected to the unit 3-ball using the two-to-one
map $\eta$, defined by 
\begin{equation}
\eta(R_{\hat{r}}(\phi))\equiv sin(\frac{\phi}{2})\hat{r}\label{eq: 4phi, the mapping of SU(2) to the unit ball}
\end{equation}
 Hence, a path in $SU(2)\times SU(2)$ can be presented as two curves
in $B\subset\mathbb{\mathbb{R}}^{3}$ by applying $\eta$ on both
its components. Figure 2 uses $\eta$ to illustrate the relations
of \emph{$U$}, $U^{RC}$ and $\tilde{U}^{RC}\equiv U^{R}(t)U(T)\otimes U(t)=U(T-t)\otimes U(t)$. 

\begin{figure}
\begin{centering}
\includegraphics[scale=0.32]{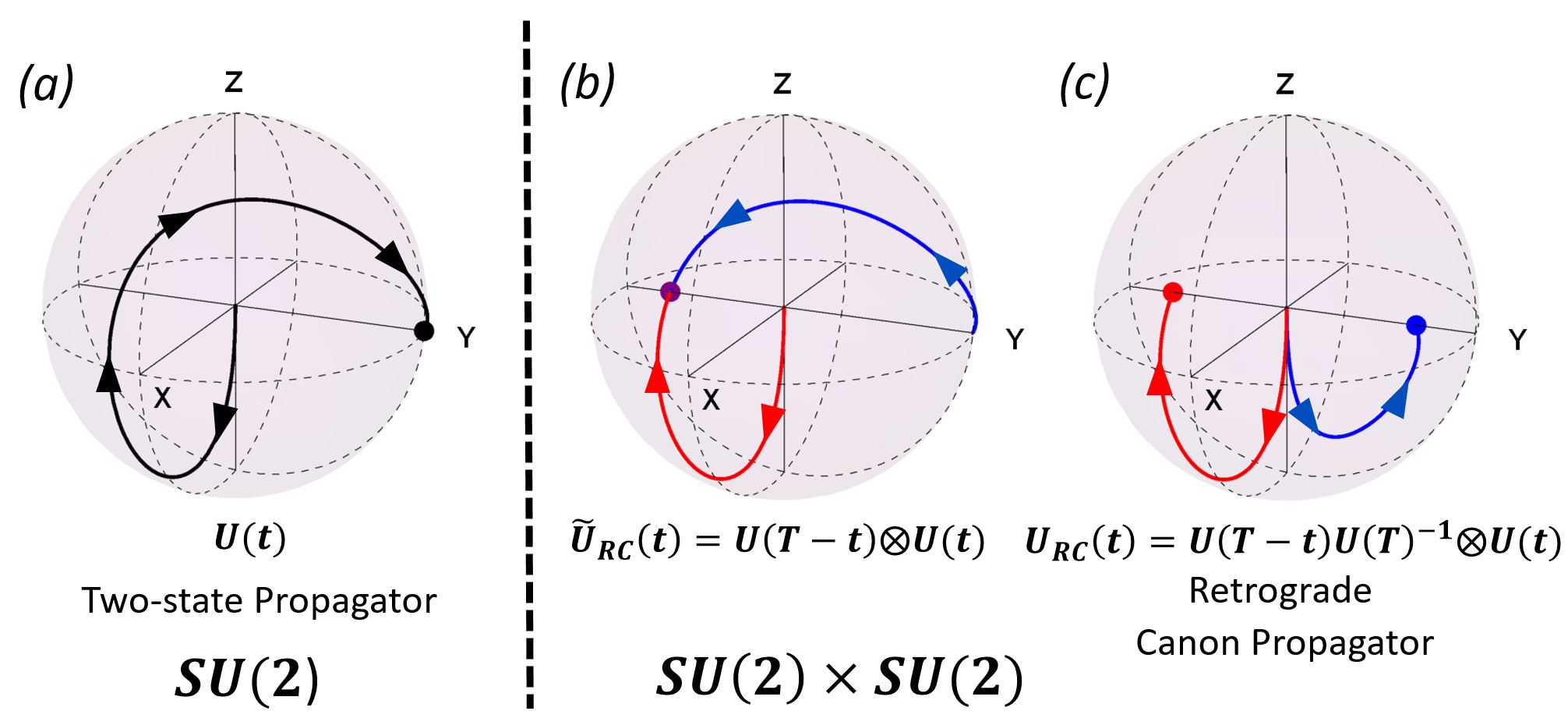}
\par\end{centering}
\centering{}\caption{\textbf{Geometrical description of the quantum retrograde canon -
}(a) The time evolution of $U:[0,T]\rightarrow SU(2)$, a two-state
propagator satisfying $U(T)=R_{\hat{y}}(\pi)$, which is mapped by
$\eta$ to the point $(0,1,0)$ - like any propagator, it starts at
the identity $I_{2}\in SU(2)$ - which is mapped by $\eta$ to the
origin \emph{$(0,0,0)$}. (b) presents $\tilde{U}^{RC}:[0,T/2]\rightarrow SU(2)\times SU(2)$,
with two curves: \emph{$\eta(U(t))$}, the curve of original propagator
(in red), and \emph{$\eta(U(T-t))$}, a curve which goes backwards
along same path (in blue). The two curves meet at time \emph{$\tau=T/2$}.
(c) presents \emph{$U^{RC}$}, the retrograde canon propagator - the
multiplication of $\tilde{U}^{RC}$ by $U(T)^{-}{}^{1}\otimes I_{2}$
transposes the blue curve\textquoteright s starting point to the origin.}
\end{figure}
 Hence, for the case where $U(T)(1,0)^{\intercal}=(0,1)^{\intercal}$,
$\eta(U(t))$ is a curve in the unit 3-ball that goes from the origin
$(0,0,0)$ to the boundary point $(0,1,0)$. The curve $\eta(U(T-t))$
goes backwards along the same path. CPT will occur at the inevitable
moment when the two curves meet, i.e., at $\tau=T/2$, and possibly
at other moments if $\eta(U(t))$ is self-intersecting. 

In our first example we examine two-state Hamiltonians of the form:
\begin{equation}
H(t)=-\frac{1}{2}\begin{cases}
p\cdot\sigma_{x} & 0\text{\ensuremath{\le}}t<\pi\\
q\cdot\sigma_{z} & \pi\text{\ensuremath{\le}}t\text{\ensuremath{\le}}2\pi
\end{cases}\label{eq:16 pyth original two level hamiltonian}
\end{equation}
where $p,q$ are odd integers. An Hamiltonian of the above form facilitates
a $p\pi$-rotation around the the $\hat{x}$-axis, followed by a $q\pi$-rotation
around the $\hat{z}$-axis. Such sequences are equivalent to a $\pm\pi$-rotation
around the $\hat{y}$-axis. Hence, these Hamiltonians facilitate CPT
and satisfy the left hand side of \ref{eq: CPT iff CPT in four} with
$|\downarrow\rangle=\pm(0,1)^{\intercal}$ for time $T\equiv2\pi$.
We shall translate $H(t)$ to a four-level system, using the the retrograde
canon and the following conjugating matrix $W$
\begin{equation}
W\equiv\frac{1}{\sqrt{2}}\begin{pmatrix}1 & \enspace sin(\theta) & 0 & \enspace cos(\theta)\\
0 & \enspace cos(\theta) & 1 & -sin(\theta)\\
0 & \enspace cos(\theta) & -1 & -sin(\theta)\\
1 & -sin(\theta) & 0 & -cos(\theta)
\end{pmatrix}\label{eq: W - more general, used in the examples}
\end{equation}
where $\theta=-\tan^{-1}(\frac{p}{q})$ . The result, $\mathcal{H^{CRC}}$,
is a constant coupling four-state Hamiltonian which facilitates CPT
from $\Psi_{1}$ to $\Psi_{3}$ at time $\tau=T/2=\pi$, where

\begin{equation}
\mathcal{H^{CRC}}=n(p,q)\cdot\begin{pmatrix}0 & C(p,q) & 0 & 0\\
\ldots & 0 & B(p,q) & 0\\
\ldots & \ldots & 0 & A(p,q)\\
\ldots & \ldots & \ldots & 0
\end{pmatrix}\label{eq:17 pyth four level Hamiltonian}
\end{equation}
where $n(p,q)=1/\sqrt{p^{2}+q^{2}}$ and $A(p,q),B(p,q),C(p,q)$ are
time-constant integers, depending on $p$ and $q$, which satisfy
the Pythagorean relation $A^{2}(p,q)+B^{2}(p,q)=C^{2}(p,q)$. The
fact that such Hamiltonians perform CPT was derived and utilized in
previous works \citep{PythagoreanPhysRevA84013414} \citep{hiddenqubit}.
Figure 3 presents the dynamics of the original two-state system and
the resulting retrograde canon system. 
\begin{figure}
\begin{centering}
\includegraphics[scale=0.45]{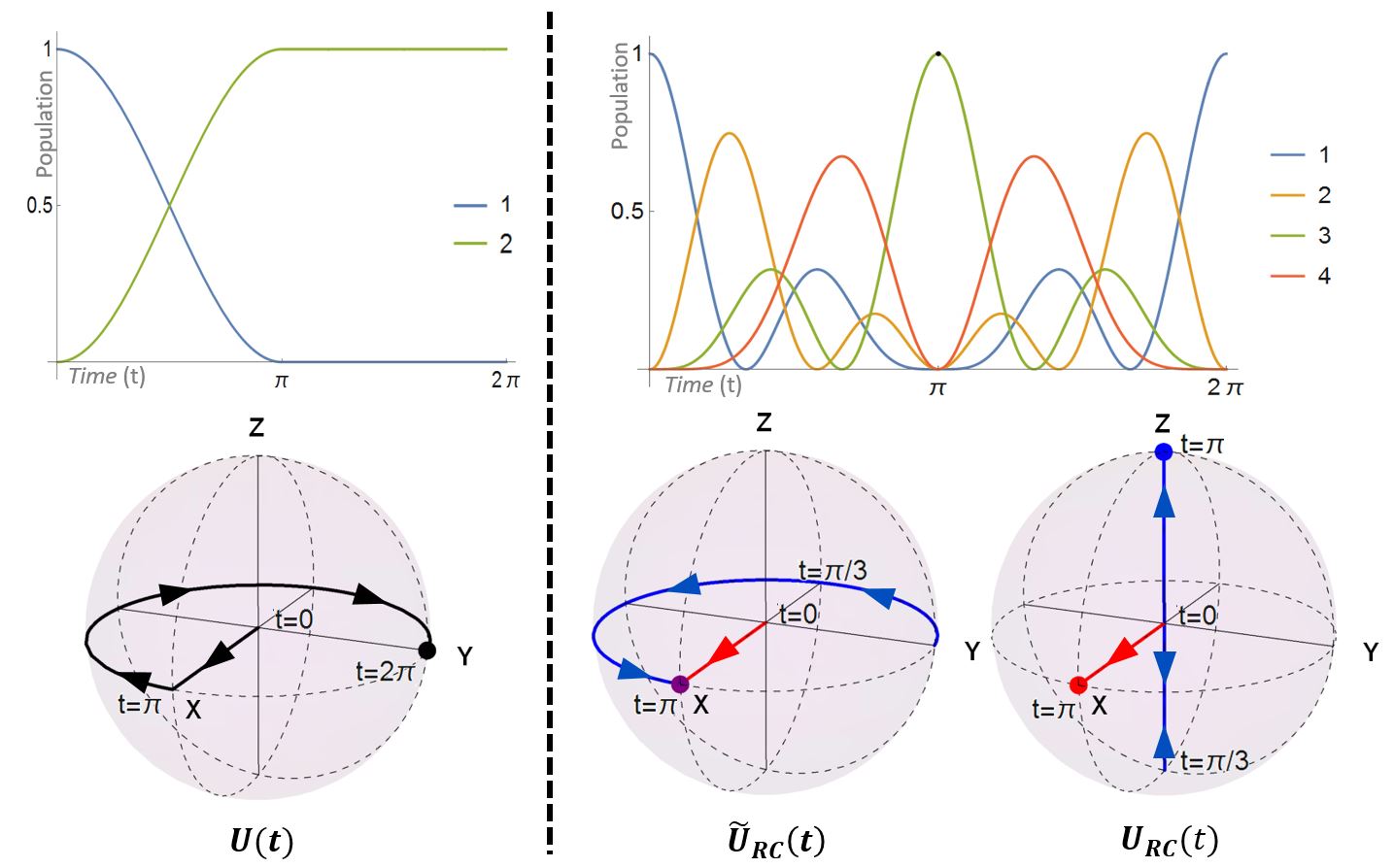}
\par\end{centering}
\caption{Pythagorean coupling CPT solutions reproduced by applying the retrograde
canon to a sequence of $\pi$-rotations. The dynamics for $p=1\enspace,q=3$
in eq. \ref{eq:16 pyth original two level hamiltonian}, which corresponds
to $(A,B,C)=(3,4,5)$ in eq. \ref{eq:17 pyth four level Hamiltonian},
are presented above. States' populations vs. interaction time appear
in the upper row while the propagator's path, visualized using $\eta$
(see eq. \ref{eq: 4phi, the mapping of SU(2) to the unit ball}),
are in the lower row. The original two-state system, on the left,
first makes a $\pi$-rotation around $\hat{x}$-axis - already reaching
CPT - and then makes a $3\pi$-rotation around $\hat{z}$-axis - which
doesn't effect the states' population. The retrograde canon operator
undergoes both rotations simultaneously on two independent qubits,
reaching CPT when both are complete. }
\end{figure}

For the second example, we take a two-state time-dependent Hamiltonian
that performs CPT through adiabatic following \citep{vitanov} by
means of a Landau-Zener scheme \citep{landau1932physik}\citep{zenerPhysRev.40.502}.
We define $H(t)\equiv\vec{h}(t)\cdot\vec{J}^{(2)}$ with 
\begin{equation}
\vec{h}(t)=(\Omega_{0},0,B(1-\frac{\ensuremath{2t}}{\tilde{T}}))\label{eq: adiabtic parameters formula}
\end{equation}
 satisfying $B\gg\Omega_{0}^{2}$ and $\Omega_{0}^{2}\gg B/\tilde{T}$
. We define the retrograde Hamiltonian with $T=\tilde{T}$ as the
moment of two-state CPT. For $t\approx T$, $U(t)(1,0)^{\intercal}\approx e^{i\alpha(t)}(0,1)^{\intercal}$,
with $\alpha(t)$ changing rapidly. Applying the method requires using
$\alpha(T)$ in the definition of $W$. Alternatively, we can rotate
$H(t)$ around the $\hat{z}$-axis with $H(t)\rightarrow R_{\hat{z}}(-\alpha(T))H(t)R_{\hat{z}}(\alpha(T)$),
and use $W$ defined in eq. \ref{eq: W - more general, used in the examples},
to get the following time-dependent Hamiltonian:
\begin{multline}
\mathcal{H}_{CRC}(t)=\\
\begin{pmatrix}0 & h_{z}(t)sin(\theta) & isin(\alpha) & h_{z}(t)cos(\theta)\\
\ldots & 0 & -cos(\alpha)sin(\theta) & 0\\
\ldots & \ldots & 0 & cos(\alpha)cos(\theta)\\
\ldots & \ldots & \ldots & 0
\end{pmatrix}\label{eq:18 fourl level adiabatic example}
\end{multline}

Figure 4 presents the resulting dynamics. It can be observed in the
plot of the original propagator's dynamics (bottom left), that for
$t\ll T/2$ the two-state propagator is approximately a rotation \emph{by
some angle} around $\hat{z}$-axis, while for $t\gg T/2$ it is approximately
a rotation by an angle $\pi$ around \emph{some axis} in the $\hat{x}$-$\hat{y}$
plain (see eq. \ref{eq: 4phi, the mapping of SU(2) to the unit ball}).
While this explains the robustness of the two-state dynamics starting
at $(1,0)^{T}$, it also explains the transient nature of the CPT
in the retrograde canon system: Using eqs. \ref{eq:1the retrograde propagator},\ref{eq:2retrograde canon operator-1}
and \ref{eq:double rotation formula} we see that if $t$ is sufficiently
far from $T/2$, then $U^{RC}(t)w_{1}\approx F(R_{\hat{z}}(\beta)R_{\hat{y}}(\pi)R_{cos(\gamma)\hat{x}+sin(\gamma)\hat{y}}(\pi))=F(R_{\hat{z}}(\delta))$
with $\beta,\delta\in[0,4\pi)$ $\gamma\in[0,2\pi)$. The low occupancy
of the third state for such $t$ is then explained by $0=\langle F(R_{\hat{z}}(\delta)),F(R_{\hat{y}}(\pi))\rangle$
$\forall\delta\in[0,4\pi)$ - where $\text{\textlangle}\cdot,\cdot\text{\textrangle}$
is the standard inner product.

\begin{figure}
\begin{centering}
\includegraphics[scale=0.39]{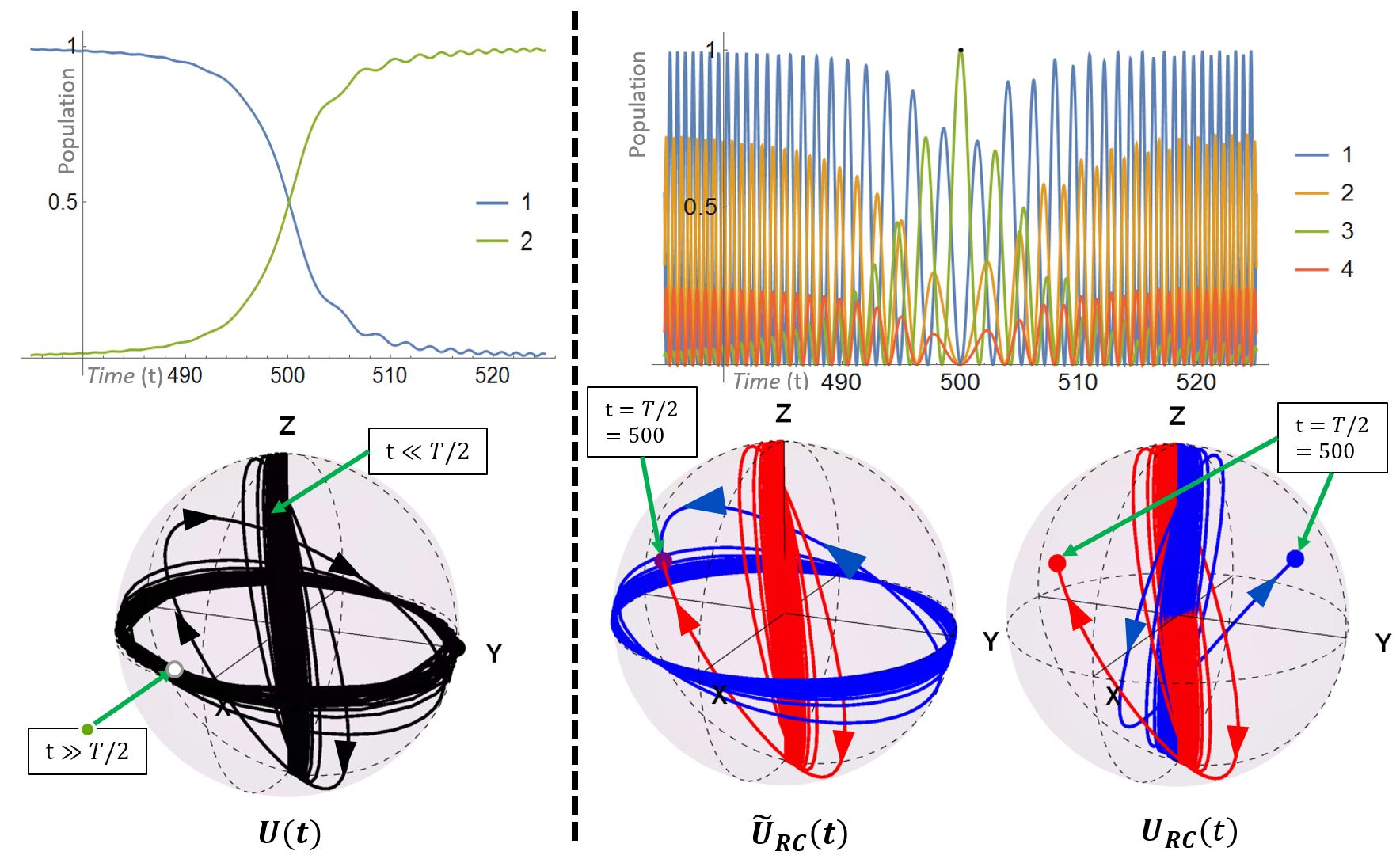}
\par\end{centering}
\centering{}\caption{Dynamics of a time-dependent four-level Hamiltonian produced by applying
the retrograde canon method to an adiabatic two-state scheme (for
$\Omega_{0}=1$, $B=100$$,\tilde{T}=1000$, in eq. \eqref{eq: adiabtic parameters formula}
and $\theta=\pi/3,$ $\alpha=1.2$ in eq. \eqref{eq:18 fourl level adiabatic example}). }
\end{figure}

\section{optimal and operator control}

Next, we indicate ways in which the retrograde canon can be used for
optimal control and operator control problems. We begin with optimal
control. It is typical for such problems to contain optimization criteria
or constraints relating to some norm defined through the parameters
of the Hamiltonian \citep{domenicodalessandro2008}. We therefore
note that a simple relation between natural norms of the four-state
$\mathcal{H}^{CRC}$ and the two-state $H$, holds under the suggested
translation method, namely
\begin{equation}
\frac{1}{\sqrt{2}}\|\mathcal{H}^{CRC}(t)\|_{F}=\sqrt{\text{\ensuremath{\|}}\vec{h}(t)\|_{2}^{2}+\|\vec{h}(2\tau-t)\|_{2}^{2}}\label{eq:14 1st relation between norms}
\end{equation}
 where $\vec{h}(t)$ is defines the two-state Hamiltonian through
$H(t)=\vec{h}(t)\cdot\vec{J}^{(2)}$, $\|\vec{h}\|_{2}$ is the Euclidean
norm and $\text{\textbardbl}\mathcal{H}\text{\textbardbl}_{F}$ is
the Frobenius norm defined by $\|\mathcal{H}\|_{F}\equiv\sqrt{\sum|\mathcal{H}{}_{ij}|^{2}}$
. It follows from eq. \ref{eq:14 1st relation between norms} that
\begin{equation}
\frac{1}{2}\intop_{0}^{\tau}\|\mathcal{H}_{CRC}(t)\|_{F}^{2}\,dt=\intop_{0}^{2\tau}\|\vec{h}(t)\|_{2}^{2}\,dt\label{eq:15 2nd relation between norms}
\end{equation}
Relations such as eqs. \ref{eq:14 1st relation between norms} and
\ref{eq:15 2nd relation between norms} provide a basis for identifying
certain multi-state optimal control problems with two-state optimal
control problems. 

Up to now we have only discussed the retrograde canon in the context
of the CPT state-control problem, yet the retrograde canon method
may also shed light on operator control problems. Indeed, knowing
that $\mathcal{U}^{CRC}$ performs CPT at time $\text{\ensuremath{\tau}}$
holds only partial information on $\mathcal{U}^{CRC}(\tau)$. However,
further information regarding $\mathcal{U}^{CRC}(\tau)$ can be related
to information on $U(\tau)$ \textendash{} i.e., the point where the
curve of $U(t)$ and $U(T-t)$ meet. Consider, for example, four-state
propagator operators of the following form $R=\text{|eg\text{\textrangle}\textlangle}ee|\pm\text{|ee\text{\textrangle}\textlangle}eg|\pm\text{|ge\text{\textrangle}\textlangle}ge|\pm\text{|gg\textrangle\textlangle}gg|$
where $|g\text{\textrangle}$,|e\textrangle{} form a basis of a two-state
system and $\text{|gg\textrangle\ensuremath{\equiv\text{|g\textrangle\ensuremath{\otimes}|g\textrangle}}}$.
Operator such as $R$ are universal operators for quantum computation,
since they maximally entangle separable states ($R\{\text{|(e+g)\ensuremath{\otimes}e\text{\textrangle}}\}=\text{|ee\text{\textrangle\ensuremath{\pm}}|gg\textrangle}$)
and thus satisfy a criterion of being a universal gate \citep{brylinski2002universal}.
If we identify $(\Psi_{1},\Psi_{2},\Psi_{3},\Psi_{4})$ with $(\uparrow\uparrow,\uparrow\downarrow,\downarrow\uparrow,\downarrow\downarrow)$,
and use the following conjugating matrix 
\begin{equation}
W\equiv\frac{1}{\sqrt{2}}\begin{pmatrix}1 & 0 & \enspace i\,sin(\theta) & \enspace i\,cos(\theta)\\
0 & -1 & \enspace i\,cos(\theta) & -i\,sin(\theta)\\
0 & 1 & \enspace i\,cos(\theta) & -i\,sin(\theta)\\
1 & 0 & -i\,sin(\theta) & -i\,cos(\theta)
\end{pmatrix}\label{eq: simlarity matrix for operator control}
\end{equation}
 then, for $\theta=0$, the condition for getting $\mathcal{U}^{CRC}(\tau)=R$
is that $\eta(U(\tau))$ is either $(0,\pm1/\sqrt{2},0)$ or $(\pm1/\sqrt{2},0,\pm1/\sqrt{2})$.
Changing $\theta$ relates the above points to other operators \textendash{}
for example, for $\theta=\pi/4$, the meeting points $(\pm1/\sqrt{2},0,\pm1/\sqrt{2})$
correspond to a \textquotedblleft double-rail\textquotedblright{}
operator, making the transitions $e_{1}\leftrightarrow e_{2}$ and
e$_{3}\leftrightarrow e_{4}$.

\section{generalizations}

Last, we present the general version of the method, which allows translating
two-level schemes to a wide family of $SU(2)\text{\texttimes}SU(2)$
controlled $n^{2}$-level systems. The generalization is based on
relinquishing three assumptions made above which concern: (a) the
pace of movement along the dynamical path traced by the original one-qubit
Hamiltonian; (b) the dimension of representation of the original Hamiltonian;
and (c) the final state of the original system. For a discussion of
these assumptions and a proof of following general translation claim
see Appendix B.

To carry out the generalization we revise our definitions. We begin
by taking a higher order representation of the original Hamiltonian:
Given a two-state Hamiltonian, $H(t)$, we define $H^{(n)}(t)$ to
be its image in an $n$-dimensional representation. That is, 
\begin{equation}
H_{n}(t)\equiv\pi_{n}(H(t))\label{eq: n dimensional rep of original ham}
\end{equation}
 where $\pi_{n}$ is a $n$-dimensional irreducible representation
of $su(2)$, fixed by satisfying $\pi_{n}(J_{i}^{(2)}$)$=J_{i}^{(n)}$
for $i=1,2,3$ where $\vec{J}^{(n)}\equiv(J_{1}^{(n)},J_{2}^{(n)},J_{3}^{(n)})$
satisfies the commutation relation $[J_{i}^{(n)},J_{j}^{(n)}]=i\varepsilon_{ijk}J_{k}^{(n)}$
with a real $J_{1}^{(n)}$ and a diagonal $J_{3}^{(n)}$. Next, we
define a $n$-dimensional retrograde Hamiltonian that goes back along
the original trajectory in a non-constant pace. That is, we define
\begin{equation}
H_{n}^{R}(t)\equiv-\dot{r}(t)H^{(n)}(T-r(t))\label{eq: 23 retrograde Hamiltonian varying pace}
\end{equation}
where $0\leq r(t)\leq T$ is a general differentiable function of
time. It can be verified by differentiation that $U_{n}^{R}(t)=U_{n}(T-r(t))U_{n}(T)^{-1}$
is the propagator generated by $H_{n}^{R}$, where $U_{n}$ is the
$n$-dimensional propagator generated by $H_{n}$. We continue be
defining the $n^{2}$\emph{-state retrograde canon Hamiltonian}, $H_{n}^{RC}$
by 
\begin{equation}
H_{n}^{RC}(t)\equiv H_{n}^{R}(t)\text{\ensuremath{\otimes}\ensuremath{\ensuremath{I_{2}}+\ensuremath{I_{2}\otimes H_{n}}(t)}}\label{eq:24 Hrc generalization}
\end{equation}
Clearly $U_{n}^{RC}$, the propagator generated by $H_{n}^{RC}$,
satisfies $U_{n}^{RC}(t)=U_{n}^{R}(t)\otimes U_{n}(t)$. 

We will also to generalize the definition of $W$, the conjugating
matrix defined above in section III: Given $n\in\mathbb{N}$ ,$k\in\{1,\ldots,n-1$\},
and a unit vector $\hat{r}\in\mathbb{R}^{3}\enspace(|\hat{r}|=1)$,
we shall designate as a $n^{2}$\emph{-conjugating matrix of $k$
and $\hat{r}$} any $n^{2}$-dimensional unitary matrix $W(k,\hat{r})=(w_{1},w_{2},...,w_{n^{2}})$
satisfying 
\begin{equation}
w_{1}\propto F_{n}(R_{\hat{r}}^{(n)}(\omega_{k})\cdot R_{\hat{y}}^{(n)}(\pi));\enspace w_{2}\propto F_{n}(R_{\hat{y}}^{(n)}(\pi))\label{eq:25 W generalization-1}
\end{equation}
where $\omega_{k}\equiv\frac{2k\pi}{n}$, $\hat{R}_{r}^{(n)}(\phi)\equiv e^{i\phi\hat{r}\cdot\vec{J}^{(n)}}$
and $F_{n}$ is the $n$-dimensional analog of the flattening function
$F$ encountered in eq. \ref{eq:double rotation formula}, i.e., it
takes a $n$-by-$n$ matrix and returns a $n^{2}$column vector defined
by $F(m)_{(n-1)i+j}\equiv m_{ij}$. 

Now, we can define the \emph{$n^{2}$-state conjugated retrograde
canon Hamiltonian}:
\begin{equation}
\mathcal{H}_{n}^{CRC}(t)\equiv W_{n}(k,\hat{r})^{\intercal}H_{n}^{RC}(t)W_{n}{(k,\hat{r})}.\label{eq: definition of the conjugated n dim retro canon Ham}
\end{equation}
 Its corresponding propagator, $\mathcal{U}_{n}^{CRC}$, of course
satisfies $\mathcal{U}_{n}^{CRC}=W_{n}(k,\hat{r})^{\intercal}U_{n}^{RC}W_{n}{(k,\hat{r})}$

Finally, we formulate the general translation claim: Let there be
$k,\:\hat{r}$, a $n^{2}$-state conjugating matrix $W_{n}(k,\hat{r})$,
a differentiable pace function $r:[0,T]\rightarrow\mathbb{{R}}$ and
a two-state system Hamiltonian $H:[0,T]\rightarrow su(2),$ whose
propagator is $U(t)$. Then, $\forall\tau\in(0,T)$ for which $U(T-r(\tau))=U(\tau)$
the following holds:
\begin{equation}
U(T)=\pm R_{\hat{r}}(\omega_{k})\Leftrightarrow1=|\text{\textlangle}e_{2},\mathcal{U}_{n}^{CRC}(\tau)e_{1}\text{\textrangle}|\label{eq:27general translatio claim}
\end{equation}
where $e_{i}\in\mathbb{C}^{n^{2}}i=1,2$ are standard basis vectors.
In particular, for any time $\tau_{x}$ for which $T-r(\tau)=\tau$,
$\mathcal{H}_{n}^{CRC}$ facilitates the transition $e_{1}\rightarrow e_{2}$
at time $\tau\in(0,T)$ if and only if $U(T)=R_{\hat{r}}(\omega_{k})$.
Note that for even $n$ we could take $k=n/2\enspace\text{\ensuremath{\in}}\mathbb{N}$
and $\hat{r}=\hat{y}$ to get $U(T)=R_{\hat{y}}^{(n)}(\pi)$ and $w_{1}\propto F_{n}(R_{\hat{y}}^{(n)}(\pi))$
for eqs. \ref{eq:27general translatio claim} and \ref{eq:25 W generalization-1}
respectively \textendash{} such a choice would be a straight forward
generalization of the four-state translation method, one which converts
two-state CPT schemes to $n^{2}$-state CPT schemes. 

\section{conclusion}

In conclusion, we introduced a novel method for translation between
multi-state control problems and two-state systems. The method provides
a new framework for importing the knowledge, tools and intuition related
to two-state systems, into multi-state research. In particular, the
method offers an exact reduction into two-state systems of multi-state
CPT problems that cannot be reduced by available methods. The idea
of the retrograde canon in control theory can be further explored
in future research: e.g., in the application of the retrograde canon
to other groups that have appropriate properties, such as $Sp(n)\rightarrow Sp(n)\times Sp(n)$
instead of $SU(2)\rightarrow SU(2)\times SU(2)$ or in using $k$
two-states CPT schemes together to generate a CPT scheme in a $\underbrace{SU(2)\times SU(2)\times...\times SU(2)}_{2k}$
controlled system. We hope that\textcolor{black}{{} the analytical reduction
of multi-state control problems to two-state systems offered by the
quantum retrograde canon may be helpful for a deeper understanding
of multi-state dynamics, and in simplifying the analysis in certain
cases of particular interest. }

\bibliographystyle{apsrev4-1}
\bibliography{retrograde_canon_paper_170321_1300}

\appendix

\section{The general form of the conjugating matrix}

We shall now present and explain the general form of the four-state
conjugating matrix $W$, and general form of $\mathcal{H}^{CRC}(t)$
derived from it. Remember that $w_{2},w_{4}\in\mathbb{C}^{4}$ can
be chosen to be any two vectors that complete $e^{i\phi_{1}}w_{1},e^{i\phi_{3}}w_{3}$
to an orthonormal basis of $\mathbb{C}^{4}$. This gives the choice
of $W$ six degrees of freedom: the angle $\alpha$ in the definition
of $|\downarrow\rangle$, the four phases of the four vectors, and
the orientation of $w_{3}$ and $w_{4}$ in the subspace orthogonal
to $span\{w_{1},w_{2}\}$. Since one degree of freedom can be regarded
as a global phase which changes nothing in the shape of $H(t)$, we
effectively have only five degrees of freedom. To slightly simplify
the presentation we set $\alpha=0$ in what follows and place $w_{3}$
in the second column of $W$. We can organize these degrees of freedom
by first defining a basic alternative $\tilde{W}=\{\tilde{w_{1}},\tilde{w_{3}},\tilde{w_{2}},\tilde{w_{4}}\}$
with 
\begin{gather*}
\tilde{w_{1}}\equiv\frac{1}{\sqrt{2}}F_{2}(I_{2})=\frac{1}{\sqrt{2}}(1,0,0,1)^{\intercal}\\
\tilde{w_{3}}\equiv-\frac{1}{\sqrt{2}}F_{2}(R_{\hat{y}}(\pi))=\frac{1}{\sqrt{2}}(0,1,-1,0)^{\intercal}\\
\tilde{w_{2}}\equiv-iF_{2}(R_{\hat{x}}(\pi))=\frac{1}{\sqrt{2}}(0,1,1,0)^{\intercal}\\
\tilde{w_{4}}\equiv-iF_{2}(R_{\hat{z}}(\pi))=\frac{1}{\sqrt{2}}(1,0,0,-1)^{\intercal}
\end{gather*}
Note that $\tilde{W}$ is just a permutation of the columns of $W$
defined in \eqref{eq: bell similarity transformation}. Next, we define
the general form of the conjugating matrix $W\equiv\{w_{1},w_{3},w_{2},w_{4}\}$
, using four variables, $\phi_{2},\phi_{3},\phi_{4},\theta\in[0,2\pi)$
to be

\begin{gather*}
w_{1}\equiv\tilde{w_{1}}\\
w_{3}\equiv e^{i\phi_{3}}\tilde{w_{3}}\\
w_{2}\equiv e^{i\phi_{3}}(cos(\theta)\tilde{w}_{3}+sin(\theta)\tilde{w}_{4})\\
w_{4}\equiv e^{i\phi_{4}}(cos(\theta)\tilde{w}_{4}-sin(\theta)\tilde{w}_{3})
\end{gather*}

to finally get the general form of $W$
\begin{equation}
W\equiv\frac{1}{\sqrt{2}}\begin{pmatrix}1 & 0 & \enspace e^{i\phi_{3}}sin(\theta) & \enspace e^{i\phi_{4}}cos(\theta)\\
0 & e^{i\phi_{2}} & \enspace e^{i\phi_{3}}cos(\theta) & -e^{i\phi_{4}}sin(\theta)\\
0 & e^{-i\phi_{2}} & \enspace e^{i\phi_{3}}cos(\theta) & -e^{i\phi_{4}}sin(\theta)\\
1 & 0 & -e^{i\phi_{3}}sin(\theta) & -e^{i\phi_{4}}cos(\theta)
\end{pmatrix}\label{eq: the general form of the conjugating matrix}
\end{equation}

We proceed to present the form of $\mathcal{H}^{CRC}$, the form of
the Hamiltonian resulting from a general two state system Hamiltonian
and a choice of $W$. The general form of $\mathcal{H}^{CRC}$ shows
which four state Hamiltonians can be translated by the retrograde
canon method to a two-level system. We parameterize the two-state
Hamiltonian by writing $H(t)=\frac{1}{2}\vec{h}(t)\cdot\vec{\sigma}$
where $\vec{h}(t)\equiv(h_{x}(t),h_{y}(t),h_{z}(t))\in\mathbb{R}^{3}$.
Then, in order to simplify the form of $\mathcal{H}^{CRC}(t)$ we
introduce the vectors $\vec{A^{\pm}}(t)\equiv(a^{\pm}(t),b^{\pm}(t),c^{\pm}(t))\equiv\frac{\vec{h}(t)\text{\textpm}\vec{h}(T-t)}{2}$
and $\vec{A}_{\theta}^{\pm}\equiv(a_{\theta}^{\pm}(t),b^{\pm}(t),c_{\theta}^{\pm}(t))$
which are defined from $\vec{A^{\pm}}(t)$ through a rotation of $\theta$
around $\hat{y}$-axis, i.e., $a_{\theta}^{\pm}(t)=cos(\theta)a^{\pm}(t)+sin(\theta)c^{\pm}(t)$
and $c_{\theta}^{\pm}(t)=cos(\theta)c^{\pm}(t)-sin(\theta)a^{\pm}(t)$.
we finally get the general matrix form of $\mathcal{H}^{CRC}(t)$,
presented in terms of the parameters of $H(t)$ and the degrees of
freedom inherited from $W$ : 
\begin{multline}
\mathcal{H}^{CRC}(t)=\\
\begin{pmatrix}0 & -ie^{i\phi_{2}}b^{+}(t) & e^{i\phi_{3}}a_{\theta}^{-}(t) & e^{i\phi_{4}}c_{\theta}^{-}(t)\\
\ldots & 0 & -e^{i(\phi_{3}-\phi_{2})}c_{\theta}^{+}(t) & e^{i(\phi_{4}-\phi_{2})}a_{\theta}^{+}(t)\\
\ldots & \ldots & 0 & ie^{i(\phi_{4}-\phi_{3})}b^{-}(t)\\
\ldots & \ldots & \ldots & 0
\end{pmatrix}\label{eq:12 Hcrc general form-1}
\end{multline}

The under-diagonal entries in eq. \ref{eq:12 Hcrc general form-1}
follow from Hermiticity. A permutation of the conjugating matrix columns
(see eq. \ref{eq: the general form of the conjugating matrix}) would
shuffle the places of the ``$a,b,c"$ letters and $"+/-$'' signs
in eq. \ref{eq:12 Hcrc general form-1}.

We note that with an appropriate choice of $\phi_{2},\phi_{3},\phi_{4}$
all six above-diagonal entries of $\mathcal{H}^{CRC}(t)$ in eq. \ref{eq:12 Hcrc general form-1}
can be set imaginary, yet no more than four can be real. There are
three possible ways to get four real above-diagonal entries - in each,
either the couple $(a_{\theta},A_{\theta})$ or $(b_{\theta},B_{\theta})$
or $(c_{\theta},C_{\theta})$ will have imaginary coefficients. Figure
2  presents an example of such couplings. Suppose we wonder whether
$\mathcal{H}(t)$, a Hamiltonian of the form in eq. \ref{eq:12 Hcrc general form-1}
performs CPT at time $\tau$. Such questions can be reduced to questions
regarding a two-state Hamiltonian $H(t)$ by applying the method backwards,
i.e., by inverting eq. \ref{eq: Hcrc definition} - while taking suitable
phase parameters $\phi_{i}\enspace(i=2,3,4)$ and choosing $\theta\in[0,2\pi)$
at will - followed by applying eq. \ref{eq: the inverse of the retrograde canon}.
The resulting two-state Hamiltonian $H(t)$ would facilitate CPT from
$|\uparrow\rangle$ to $|\downarrow\rangle$ at time $2\tau$ if and
only if the four-state Hamiltonian $\mathcal{H}(t)$ would evolve
the state $e_{1}$ to $\pm e_{2}$ at time $\tau$. 

\section{The general translation claim - discussion and proof}

The general version of the retrograde canon method allows translating
two-level schemes to a wide family of $SU(2)\times SU(2)$ controlled
$n^{2}$-level systems. The definitions of the general method and
its corresponding general translation claim appear in eqs. \eqref{eq:24 Hrc generalization}-\eqref{eq:27general translatio claim}
in the main text. The first part of this appendix is concerned with
explaining the rational behind the generalization, while the second
part contains a proof of the general translation claim. 

\subsection{\label{app:subsec}A discussion of the generalizations}

We recall that the method's general version is based on relinquishing
three assumptions that underlie the fundamental version - assumptions
which concern: (a) the pace of movement along the path of the original
propagator; (b) the dimension of representation of the original Hamiltonian;
and (c) the final state of the original system. Let us explain the
rational of dropping these three assumptions. 

We begin with (a), the pace of movement along the path of the original
propagator. The definition of $H^{RC}$, given in eq. \eqref{eq: 23 retrograde Hamiltonian varying pace},
is such that $U(t)$ and $U^{R}(t)$ move at the same constant pace
(albeit in opposite directions). This, however, is not a necessary
condition for the method to work. The propagator, $U^{R}$ can move
at practically any pace, $\dot{r}(t)$, so long as there's a moment
$\tau$ for which $U(T-r(\tau))=U(\tau)$ \textendash{} which always
happens for a time $\tau_{r}\in[0,T]$ such that $T-r(\tau_{r})=\tau_{r}$.
Thus, by loosening the pace assumption allows creating a wide family
of significantly different multi-state Hamiltonians, even if the pace
is constant (i.e. $r(t)=at$ for some $a>0$). Changing the pace can
be used to move $U(\tau_{r})$ - the meeting point of $U(t)$ and
$U(T-t)$ - which defines the operator $\mathcal{U}^{CRC}(\tau_{r})$,
to any point along the path of $U(t)$. 

Next, consider assumption (b), concerning the dimension of representation
of the original Hamiltonian: We shall see in the proof of the general
translation claim below that in the definition of the retrograde canon
Hamiltonian we need not restrict ourselves to using the fundamental
representation of the original Hamiltonian. Correspondingly, the output
system does not have to be a four-level system. Rather, it can be
a $n^{2}$-state system for every $n\in\mathbb{N}$. The change of
dimension of the retrograde canon Hamiltonian has to be accompanied
by a non-trivial revision of the definition of the conjugating matrix
$W$. One natural way of revising $W$ - which is appropriate only
for \emph{even} $n$ - is defining the $n^{2}$-state conjugating
matrix as any $n^{2}$-dimensional unitary matrix $W_{n}=\{w_{1},w_{2},...,w_{n^{2}}\}$
such that 

\[
w_{1}\propto F_{n}(I_{n});\enspace w_{2}\propto F_{n}(R_{\hat{y}}^{(n)}(\pi))
\]
where $F_{n}$ is the flattening function presented in \eqref{eq:25 W generalization-1}.
Note, that while $I_{n}$ has $1$s on the diagonal, and $0$s everywhere
else, the matrix $R_{\hat{y}}^{(n)}(\pi)$ has $1$s and $(-1)$s
alternately on the anti-diagonal, and $0$s everywhere else. Therefore,
since for even $n$ the anti-diagonal and the diagonal of a $n$-dimensional
matrix have no common entry, $w_{1}$ and $w_{2}$ will indeed by
orthogonal for every even $n$. On the other hand, for odd $n$, the
anti-diagonal and the diagonal do have a common entry, and therefore
$F_{n}(I_{n})$ and $F_{n}(R_{\hat{y}}^{(n)}(\pi))$ will not be orthogonal
and cannot be columns of the same unitary matrix. 

The generalization of the third assumption (c), regarding the final
state of the original system $U(T)$, is designed to solve the above
mentioned problem of odd $n$ representations. In the process, it
opens up the method to a wider range of conjugating matrices and two-state
schemes, thus enabling the production of a wider variety of $n^{2}$-state
CPT schemes \textendash{} useful also for even $n>2$. The generalization
with regards the final state of the original system, $U(T)$, comes
from the insight that the method and the translation claim essentially
rely on just three basic conditions (assuming for the moment that
$\alpha=0$). To present these conditions we mark the first two columns
of the required $n^{2}$-state conjugating matrix as $w_{1}\equiv F_{n}(m_{1})$
and $w_{2}\equiv F_{n}(m_{2})$. 

The first condition is that 
\begin{equation}
U_{n}(T)^{-1}m_{1}=m_{2}\label{eq: first of three conditions}
\end{equation}

The second condition is that 
\begin{equation}
m_{2}=R_{\hat{y}}^{(n)}(\pi)\label{eq: second of three conditions}
\end{equation}

The third condition is that 
\begin{equation}
F_{n}(m_{1})\perp F_{n}(m_{2})\label{eq:third of three conditions}
\end{equation}
 The last condition simply ensures that $w_{1}$ and $w_{2}$ are
orthogonal and can therefore be two columns in the same unitary matrix.
The role of the first two conditions shall be clarified in the proof
below. If these conditions are satisfied we can define as the $n^{2}$-state
conjugating matrix any unitary matrix whose first two columns are
$F_{n}(m_{1})$ and $F_{n}(m_{2})$, and formulate a general translation
claim for two-state systems whose final states satisfy \eqref{eq: first of three conditions},
i.e. $U_{n}(T)=m_{2}m_{1}^{-1}$.

Interestingly, assuming eqs. \eqref{eq: first of three conditions}
and \eqref{eq: second of three conditions}, all that is needed to
satisfy eq. \eqref{eq:third of three conditions} is that $U(T)$,
the original propagator at time $T$, should satisfy 
\begin{equation}
\chi_{n}(U(T))=0\label{eq: character equals zero}
\end{equation}
 Where $\chi_{n}:SU(2)\rightarrow\mathbb{C}$ is the character of
the $n$-dimensional irreducible representation of $SU(2)$ ,i.e.,
the function which for every element of $SU(2)$ returns the trace
of its image in the $n$-dimensional irreducible representation. Hence,
\eqref{eq: character equals zero} is equivalent to 
\begin{equation}
trace(U_{n}(T))=0\label{eq: trace equals zero}
\end{equation}
 We shall now prove that indeed, if eqs. \eqref{eq: first of three conditions}
and \eqref{eq: second of three conditions} are satisfied then eq.
\eqref{eq: trace equals zero} follows from \eqref{eq:third of three conditions}.
We mark $Y_{n}=R_{\hat{y}}^{(n)}(\pi)$. Now, from \eqref{eq: first of three conditions}
and \eqref{eq: second of three conditions} we get $m_{2}\equiv\pm Y_{n}$
and $m_{1}\equiv U_{n}(T)Y_{n}$. Therefore,
\begin{equation}
\langle F_{n}(m_{1}),F_{n}(m_{2})\rangle=\langle F_{n}(U_{n}(T)Y_{n}),F_{n}(Y_{n})\rangle\label{eq: what orthogonality mean}
\end{equation}
Note that for every $A,B\in M_{n}(C)$ the following identity holds
\begin{equation}
\langle F_{n}(A),F_{n}(B)\rangle=\langle F_{n}(AY_{n}),F_{n}(BY_{n})\rangle\label{eq: inner product perserved by multiplication from right}
\end{equation}
Eq. \eqref{eq: inner product perserved by multiplication from right}
can be understood as following from the form of $Y_{n},$ whose only
non-zero entries are $1$s and $(-1)$s on the anti-diagonal. Therefore,
multiplication by $Y_{n}$ from the right simply permutes the columns
of the multiplied matrix while providing factors of $\pm1$. Hence,
applying $Y_{n}$ to the matrices on both sides of the inner product
only changes the order of summation and not the result. Using \eqref{eq: inner product perserved by multiplication from right},
and noting the fact that $Y_{n}Y_{n}=-I_{n}$, we see that under assumptions
\eqref{eq: first of three conditions} and \eqref{eq: second of three conditions},
eq. \eqref{eq:third of three conditions} is indeed equivalent to
\eqref{eq: character equals zero}, since

\begin{multline}
0=\text{\textlangle}F_{n}(m_{1}),F_{n}(m_{2})\text{\textrangle}=\text{\textlangle}F_{n}(U_{n}(T)Y_{n}),F_{n}(Y_{n})\text{\textrangle}\\
=\text{\textlangle}F_{n}(U_{n}(T)),F_{n}(I_{n})\text{\textrangle}=trace(U_{n}(T))=\chi_{n}(U(T))\label{eq: trace equals zero final equation}
\end{multline}
 We shall now present and prove a criterion for a matrix $U(T)\in SU(2)$
to satisfy eq. \eqref{eq: character equals zero}: For every unit
vector $\hat{r}\in\mathbb{R}^{3}$ the following equivalence holds
\begin{equation}
\chi_{n}(R_{\hat{r}}(\omega))=0\Leftrightarrow\omega=2k\pi/n\label{eq: criterion for character equals zero}
\end{equation}
 $k\in Z,k\neq0$. To prove \eqref{eq: criterion for character equals zero}
we note that for every unit vector $\hat{r}\in\mathbb{R}^{3}$ there
exists, $\tilde{\pi}_{n}:SU(2)\rightarrow M_{n}[\mathbb{{C}}${]},
an irreducible $n$-dimensional representation of $SU(2)$, for which
$\tilde{R}_{\hat{r}}^{(n)}(\omega)\equiv\tilde{\pi}_{n}(R_{\hat{r}}(\omega))$
is diagonal. In such a representation we have 
\begin{equation}
\tilde{R}_{\hat{r}}^{(n)}(\omega)=\begin{pmatrix}e^{ij\omega}\\
 & e^{i(j-1)\omega}\\
 &  & \ddots\\
 &  &  & e^{-ij\omega}
\end{pmatrix}\label{eq: to diagonal from higher rep}
\end{equation}
where $j=(n-1)/2$. Therefore 
\begin{equation}
trace(\tilde{R}_{\hat{r}}^{(n)}(\omega))=\sum_{m=1-n}^{n-1}e^{im\omega/2}=\begin{cases}
\frac{sin(n\omega/2)}{sin(\omega/2)} & \omega\neq0\\
n & \omega=0
\end{cases}\label{eq: the sum of trace equation}
\end{equation}
from which \eqref{eq: criterion for character equals zero} follows.
To summarize the discussion of assumption (c), regarding the final
state of the original system, we see that conditions \eqref{eq: first of three conditions}-\eqref{eq:third of three conditions}
entail that $U_{n}(T)=(R_{\hat{r}}(2k\pi/n))$ for $k\in Z,k\neq0$,
and ensure that the definition of the $n^{2}$-state conjugating matrix
given in eq. \eqref{eq:25 W generalization-1} can be satisfied, since
the first two columns are orthogonal. The fact that under definition
\eqref{eq:25 W generalization-1}, the general translation claim,
given in eq. \eqref{eq:27general translatio claim}, follows, is what
we shall now prove. 

\subsection{\label{app:subsec-1}A proof of the general translation claim}

We need a some more preparation before presenting the proof. We shall
use the fact that a high order representation of a propagator is a
propagator of the high order Hamiltonian, i.e., that for all $t\in[0,T]$
we have 
\begin{equation}
\Pi_{n}(U(t))=U_{n}(t)\label{eq:higher order rep of propagator is the prop of higher order Hamiltonian}
\end{equation}
where $\Pi_{n}:SU(2)\rightarrow M_{n}(\mathbb{C})$ is the lie group
representation of $SU(2)$ which for every $R_{\hat{r}}(\phi)\in SU(2)$
satisfies 
\begin{equation}
\Pi_{n}(R_{\hat{r}}(\phi))\equiv R_{\hat{r}}^{(n)}(\phi)\label{eq: the definition of the Group representation}
\end{equation}
Eq. \eqref{eq:higher order rep of propagator is the prop of higher order Hamiltonian}
follows directly from the fact that $H_{n}(t)\equiv\pi_{n}(H(t))$
where $\pi_{n}:su(2)\rightarrow M_{n}[\mathbb{C}]$, is the $n$-dimensional
lie algebra linear representation of $su(2)$, fixed by $\pi_{n}(J_{i}^{(2)})=J_{i}^{(n)}$
for $i=1,2,3$. This can be proved by writing $U_{n}(t)=R_{\hat{n(t)}}^{(n)}(\phi_{n}(t))$
where $\phi_{n}:[0,T]\rightarrow\mathbb{R}$ and $\hat{r}_{n}:[0,T]\rightarrow\mathbb{S}^{2}\subset\mathbb{R}^{3}$,
and showing that for every $n\ge2$, $\phi_{n}=\phi_{2}$ and $\hat{r}_{n}=\hat{r}_{2}$.
Indeed, from \eqref{eq:higher order rep of propagator is the prop of higher order Hamiltonian}
and \eqref{eq: the definition of the Group representation} it follows
that the left side of eq. \eqref{eq:27general translatio claim} means
that $U_{n}(T)$ satisfies 
\begin{equation}
U_{n}(T)=R_{\hat{r}}^{(n)}(\omega_{k})\label{eq: how the hgiher order prop looks}
\end{equation}

Another important fact for the proof is that the analog of eqs. \eqref{eq:double rotation formula}
and \eqref{eq: the defining equation of symplectic} also holds for
higher dimensional irreducible representations of $SU(2)$ - i.e.,
that for every $u_{n}\in\Pi_{n}(SU(2))$ the following equation holds
\begin{equation}
Y_{n}=u_{n}Y_{n}u_{n}{}^{\intercal}\label{eq: Y^n=00003DuY^nu in higher dimen}
\end{equation}

We shall prove \eqref{eq: Y^n=00003DuY^nu in higher dimen} using
the defining property of group representations, which is that the
multiplication of images of group elements under the representation
equals the image of the multiplication of the group elements. Hence,
together with eqs. \eqref{eq: the defining equation of symplectic},
\eqref{eq: the definition of the Group representation}, and the fact
that there exists $u\in SU(2)$ such that $u_{n}=\Pi_{n}(u)$ we get
\begin{multline}
Y_{n}=\text{\textgreek{P}}_{n}(Y)=\text{\textgreek{P}}_{n}(uYu^{\intercal})=\\
\text{\textgreek{P}}_{n}(u)\text{\textgreek{P}}_{n}(Y)\text{\textgreek{P}}_{n}(u^{\intercal})=u_{n}Y_{n}u_{n}{}^{\intercal}\label{eq:line proof of Y=00003DuYu high order rep}
\end{multline}
we have assumed in eq. \eqref{eq:line proof of Y=00003DuYu high order rep}
that $u^{\intercal}\in SU(2)$ and that $\Pi_{n}(u^{\intercal})=(\Pi_{n}(u))^{\intercal}$
\textendash{} which are facts that can be proved, for instance, using
Euler decomposition of $u\in SU(2)$ as $u=R_{\hat{y}}(\alpha)R_{\hat{z}}(\beta)R_{\hat{y}}(\gamma)$. 

After these preliminaries we can proceed to prove the general translation
claim, given in eq. \eqref{eq:27general translatio claim}. We begin
with the $\Rightarrow$ direction of eq. \eqref{eq:27general translatio claim}.
We shall use the assumptions that $U(T-r(\tau))=U(\tau)$ and that
$U(T)=\pm R_{\hat{r}}(\omega_{k})$ and conclude that $1=|\langle e_{2},U_{n}^{CRC}(\tau)e_{1}\rangle|$.
First, note that$\langle e_{2},U_{n}^{CRC}(\tau)e_{1}\rangle=\langle w_{2},U_{n}^{RC}(\tau)w_{1}\rangle$.
From eq. \eqref{eq:higher order rep of propagator is the prop of higher order Hamiltonian}
it follows that 
\begin{equation}
U_{n}(T-r(\tau))=U_{n}(\tau)\label{eq: the assumption in higher rep}
\end{equation}
We will proceed to show that 
\begin{equation}
w_{2}=\pm U_{n}^{RC}(\tau)w_{1}\label{eq:w2=00003DUrc*w1}
\end{equation}
 from which the conclusion follows. from eqs. \eqref{eq: how the hgiher order prop looks}
and \eqref{eq: the assumption in higher rep} we see that 
\begin{equation}
U_{n}^{RC}(\tau)w_{1}=(U_{n}(\tau)R_{\hat{r}}^{(n)}(\omega_{k})^{-1}\otimes U_{n}(\tau))w_{1}\label{eq: the form of prop at meeting point}
\end{equation}
from which, using eqs. \eqref{eq:25 W generalization-1}, the generalization
of \eqref{eq:double rotation formula} and \eqref{eq: Y^n=00003DuY^nu in higher dimen},
we get
\begin{multline}
U_{n}^{RC}(\tau)w_{1}=\\
F_{n}(U_{n}(\tau)R_{\hat{r}}^{(n)}(\omega_{k}){}^{-1}R_{\hat{r}}^{(n)}(\omega_{k})Y_{n}U_{n}(\tau){}^{\intercal})\\
=\pm F_{n}(U_{n}(\tau)Y_{n}U_{n}(\tau)^{\intercal})\\
=\pm F_{n}(Y_{n})=\pm w_{2}\label{eq: getting Uw1=00003DUYU^t}
\end{multline}

For the $\Leftarrow$ direction of eq. \eqref{eq:27general translatio claim}
we shall assume that $U(T-r(\tau))=U(\tau)$ and that $1=|\langle e_{2},U_{CRC}^{(n)}(\tau)e_{1}\rangle|$
to conclude that $U(T)=R_{\hat{r}}(\omega_{k})$. From the assumption
it follows that 
\begin{equation}
e^{i\alpha}e_{2}=U_{n}^{CRC}(\tau)e_{1}\label{eq: UCRC TAKES E1 TO e2 WITH PHASE IN HIGH REP}
\end{equation}
 for some $\alpha\in[0,2\pi)$. Or equivalently, 
\begin{equation}
e^{i\alpha}w_{2}=U_{n}^{RC}(\tau)w_{1}=(U_{n}^{R}(\tau)\otimes U_{n}(\tau))w_{1}\label{eq: assumption in terms of wi in high rep}
\end{equation}
Hence, marking $\tilde{R}\equiv R_{\hat{r}}^{(n)}(\omega_{k})Y_{n}$
and $V\equiv U_{n}^{R}(\tau)\tilde{R}Y_{n}U_{n}^{-1}(\tau)$ and noting
that $Y_{n}^{-1}=\pm Y_{n}$ (where the sign depends on whether $n$
is even) we get that
\[
\begin{array}{c}
e^{i\tilde{\alpha}}Y_{n}=F^{-1}((U_{n}^{R}(\tau)\otimes U_{n}(\tau))F(\tilde{R}))=\\
=U_{n}^{R}(\tau)\tilde{R}U_{n}(\tau)=\pm VU_{n}(\tau)Y_{n}U_{n}^{\intercal}(\tau)=\pm VY_{n}
\end{array}
\]

where $\tilde{\alpha}\in[0,2\pi).$ we get 
\begin{equation}
e^{i\tilde{\alpha}}Y^{(n)}=a^{(n)}Y^{(n)}\label{eq: phase Y=00003D AY high rep}
\end{equation}
 Hence $V=\pm e^{i\tilde{\alpha}}I_{n}$. Since $V\in\Pi_{n}(SU(2))\subset SU(n)$
it follows that $1=|V|=|\pm e^{i\tilde{\alpha}}I_{n}|=(\pm1)^{n}e^{in\tilde{\alpha}}\Rightarrow$
$\tilde{\alpha}=2\pi k/n$ with $k\in\{0,1,\text{\dots}n-1\}$. hence
$V=(\pm1)^{n}e^{i2\pi k/n}I_{n}$. Yet an irreducible representations
of $SU(2)$ may contain only $I_{n}$ and possibly $-I_{n}$. Hence,
we finally get $V=\pm I_{n}$ and $U_{n}^{R}(\tau)=U_{n}(\tau)Y_{n}\tilde{R}^{-1}.$
From which it follows that 
\[
\begin{array}{c}
U(T)=(U_{n}^{R}(\tau))^{-1}U_{n}(\tau)=\pm\tilde{R}Y_{n}(U_{n}(\tau))^{-1}U_{n}(\tau)=\\
=\pm\tilde{R}Y_{n}=\pm R_{\hat{r}}^{(n)}(\omega_{k})
\end{array}
\]

\end{document}